\documentclass[aps,preprint,floats,psfig]{revtex4}
\usepackage{graphics}

\begin{document}

\title{FREQUENCY AND PHASE NOISE IN NON-LINEAR 
MICROWAVE OSCILLATOR CIRCUITS}
                              
\author{C. Tannous}
\email{tannous@univ-brest.fr}
\affiliation{Laboratoire de Magn\'etisme de Bretagne, CNRS-UMR 6135,
Universit\'e de Bretagne Occidentale, BP: 809 Brest CEDEX, 29285 FRANCE}

\date{May 16, 2003}

\begin{abstract}

We  have  developed  a  new methodology  and  a  time-domain
software  package  for  the estimation  of  the  oscillation
frequency  and the phase noise spectrum of non-linear  noisy
microwave  circuits based on the direct integration  of  the
system of stochastic differential equations representing the
circuit.
Our theoretical evaluations can be used in order to make
detailed comparisons with the experimental
measurements of phase noise spectra in selected  oscillating
circuits.


\end{abstract}

\maketitle

\section{Introduction}
Electronically   tuned   microwave   oscillators   are   key
components   used   in   a   wide   variety   of   microwave
communications systems \cite{gunger}. The phase of the  output  signal
exhibits  fluctuations  in  time  about  the  steady   state
oscillations  giving rise to Phase Noise  a  very  important
characteristic  that  influences  the  overall   performance
especially at higher microwave frequencies.

In  order  to  understand the oscillator phase behaviour,  a
statistical  model for a non-linear oscillating circuit  has
to be developed and presently, no accurate theoretical model
for phase noise characterization is available because of the
particularly difficult nature of this problem. This  is  due
to  the  hybrid  nature  of non-linear microwave  oscillator
circuits  where distributed elements (pertaining usually  to
the associated feeding or resonator circuits) and non-linear
elements  (pertaining  usually to the  amplifiying  circuit)
have to be dealt with simultaneously \cite{gunger} .

The  main  aim of this report is to establish a  theoretical
framework  for  dealing  with the  noise  sources  and  non-
linearities  present in these oscillators, introduce  a  new
methodology   to  calculate  the  resonance  frequency   and
evaluate the time responses (waveforms) for various voltages
and  currents  in  the circuit without  or  with  the  noise
present.  Once this is established, the phase noise spectrum
is determined and afterwards the validity range of the model
is  experimentally gauged with the use of different types of
microwave oscillators \cite{heinen,anzill}.
This  report  is organised in the following way:  Section  II
covers the theoretical analysis for the oscillating circuit,
reviews noise source models and earlier approches. Section III
presents  results of the theoretical analysis and highlights
the  determination  of  the  resonance  frequency  for  some
oscillator circuits without noise. In section IV, phase noise
spectra  are determined for several oscillator circuits  and
section  V  contains  the experimental results. The Appendix 
contains  circuit diagrams and corresponding state equations
for several non-linear oscillator circuits.

\section{THEORETICAL ANALYSIS}

In standard microwave analysis, it is difficult to deal with
distributed  elements in the time domain  and  difficult  to
deal  with non-linear elements in the frequency domain. Non-
linear microwave oscillator circuits have simultaneously non-
linear  elements  in  the amplifying  part  and  distributed
elements  in  the resonating part [Non-linearity  is  needed
since  it  is well known that only non-linear circuits  have
stable oscillations].

Before we tackle, in detail, the determination of the  phase
noise,  let  us describe the standard procedure for  dealing
with  the determination of resonance frequency of non-linear
oscillator circuits:

\begin{itemize}

\item The  first step is to develop a circuit model for  the
oscillator  device and the tuning elements.  The  equivalent
circuit  should  contain inherently noiseless  elements  and
noise sources that can be added at will  in various parts of
the circuit. This separation is useful for pinpointing later
on the precise noise source location and its origin \cite{heinen}. The
resulting  circuit  is described by a set  of  coupled  non-
linear differential equations that have to be written  in  a
way  such  that a linear sub-circuit (usually the resonating
part)  is coupled to another non-linear sub-circuit (usually
the oscillating part).
\item The  determination of the periodic response of the  non-
linear circuit.
\item The  third  step  entails  performing  small  signal  ac
analysis  (linearization  procedure)  around  the  operating
point.  The  result of the ac analysis is  a  system  matrix
which  is  ill-conditioned  since  a  large  discrepency  of
frequencies are present simultaneously (One has a factor  of
one  million  in  going  from kHz to GHz  frequencies).  The
eigenvalues of this matrix have to be calculated with  extra
care  due to the sensitivity of the matrix elements  to  any
numerical roundoff \cite{anzill}.

\end{itemize}

We  differ  from the above analysis, by integrating  
the state equations directly with standard/non-
standard  Runge-Kutta methods adapted to the non-stiff/stiff
system  of  ordinary differential equations.  The  resonance
frequency is evaluated directly from the waveforms  and  the
noise  is  included  at various points  in  the  circuit  as
Johnson or Shot noise.

This  allows us to deal exclusively with time domain methods
for  the noiseless/noisy non-linear elements as well as  the
distributed elements. The latter are dealt with  through  an
equivalence to lumped elements at a particular frequency.

As  far as point 3 is concerned, the linearization procedure
method  is  valid only for small-signal analysis whereas  in
this situation, we are dealing with the large signal case.

Previously, several methods have been developed in order  to
find  the  periodic  response.  The  most  well  established
methods  are the Harmonic balance and the piecewise Harmonic
balance  methods \cite{nakhla}. Schwab \cite{schwab} has combined
the time-domain (for the non-linear amplifier part) with the
frequency  domain (for the linear resonating  part)  methods
and  transformed  the system of equations  into  a  boundary
value  problem  that  yields the periodic  response  of  the
system.

\section{TIME  RESPONSES OF NON-LINEAR OSCILLATORS AND  RESONANCE
FREQUENCY DETERMINATION}

For  illustration and validation of the method  we  solve  6
different  oscillator  circuits  (The Appendix contains  the
circuit diagrams and the corresponding state equations):

\begin{itemize}
\item The standard Van der Pol oscillator.
\item The amplitude controlled Van der Pol oscillator.
\item The Clapp oscillator.
\item The Colpitts oscillator.
\item Model I oscillator.
\item Model II oscillator.
\end{itemize}

We  display  the  time  responses  (waveforms)  for  various
voltages  and currents in the attached figures for  each  of
the  six oscillators. All oscillators reach periodic  steady
state  almost instantly except the amplitude controlled  Van
der Pol (ACVDP) and the Colpitts circuits. For instance,  we
need,  typically, several thousand time steps to  drive  the
ACVDP  circuit  into  the oscillatory steady  state  whereas
several hundred thousand steps are required for the Colpitts
circuit.
Typically,  the  rest  of the circuits studied  reached  the
periodic  steady  state in only less  a  couple  of  hundred
steps.

\section{PHASE NOISE SPECTRUM EVALUATION}

Once the oscillating frequency is obtained, device noise  is
turned  on and its effect on the oscillator phase noise  is
evaluated. All the above analysis is performed  with
time domain simulation techniques.

Finally,   Fourier  analysis  is  applied  to  the  waveform
obtained  in  order  to  extract the  power  spectrum  as  a
function  of frequency. Very long simulation times  (on  the
order  of several hundred thousand cycles) are needed  since
one  expects inverse power-law dependencies on the frequency
\cite{gunger}.

We use a special Stochastic time integration method namely
the 2S-2O-2G Runge-Kutta method developed by Klauder and Peterson,
and we calculate the PSD (Power Spectral Density) from the time
series obtained.

It is worth mentioning that our methodology is valid for any
type  of  oscillator  circuit and  for  any  type  of  noise
(Additive  White  as  it is in Johnson noise  of  resistors,
Mutiplicative and Colored or $1/f^{\alpha}$ with $\alpha$ 
arbitrary as it  is for Shot noise stemming from junctions 
or imperfections inside the device).
In  addition, the approach we develop is independent of  the
magnitude of the noise. Regardless of the noise intensity we
evaluate  the time response and later on the power  spectrum
without performing any perturbative development whatsoever.
Recently,  Kartner \cite{kartner} developed  a  perturbative  approach  to
evaluate the power spectrum without having to integrate  the
state  equations. His approach is valid for weak noise  only
and  is  based  on an analytical expression  for  the  power
spectrum. Nevertheless one needs to evaluate numerically one
Fourier coefficient $g_{1,0}$ the spectrum depends on.

\section{EXPERIMENTAL VERIFICATION}

Microwave oscillators are realised using a very wide variety
of circuit configurations and resonators. We plan to design,
fabricate  and test microstrip oscillators with GaAs  MESFET
devices  with  coupled lines and ring  resonators  \cite{curtice}.  The
measured  phase noise of these oscillators will be  compared
with the theoretical prediction from the above analysis.  We
also  plan  to  apply the above analysis to the experimental
phase  results  obtained from various  electronically  tuned
oscillators  that  have  been  already  published   in   the
literature \cite{gunger,heinen,anzill,schwab}. \\

{\bf Acknowledgments}:
The author would like to thank FX Kartner and W. Anzill for sending several
papers, reports and a thesis that were crucial for the present
investigation. Thanks also to S.O. Faried who made several circuit
drawings and S. Kumar for suggesting two additional circuits (Model I and
II) to test the software.

\vspace{2cm}

\centerline{\Large\bf APPENDIX}
                                                            
\subsection{Van Der Pol oscillator}
 
{\bf State-Space Equations of Van der Pol oscillator}:\\

\begin{eqnarray}
\frac{di_L}{dt} &=& \frac{V_c}{L}\\
\frac{dV_c}{dt} &=& \frac{1}{C} (-\mu V_{c} \hspace{1mm} {i_L}^2  -i_{L} + \mu V_{c})
\end{eqnarray}

Define:\\

\begin{eqnarray}
\left[ \begin{array}{c}
X_{1} \\
X_{2}
\end{array} \right] \stackrel{Def.}{=} \left[ \begin{array}{c}
i_{L}\\
V_{c}
\end{array} \right]
\end{eqnarray}

Rewrite Equations 1 and 2 in state-space form:

\begin{eqnarray}
{\dot{X}}_{1} &=& X_{2}\\
{\dot{X}}_{2} &=& \mu X_{2} ( 1- X_{1}^2)- X_{1}
\end{eqnarray}

\subsection{Amplitude controlled Van Der Pol oscillator}

{\bf State-Space equations of Amplitude controlled Van der Pol oscillator}:\\

\begin{eqnarray}
\frac{dV_c}{dt} &=& -\frac{1}{C} (i_{L} + Gu -G_{a}(z)u-i_{r})\\
\frac{di_L}{dt} &=& \frac{u}{L}\\
\frac{dz}{dt} &=& \frac{1}{T} [ {(\frac{u}{V_{0}})}^2- z]
\end{eqnarray}

where $G_{a}(z)=G(1-c(z-1))$ is   the   voltage  controlled  conductance. \\

Choosing   the variables:

\begin{eqnarray}
\left[ \begin{array}{c}
X_{1} \\
X_{2}  \\
X_{3}
\end{array} \right] \stackrel{Def.}{=} \left[ \begin{array}{c}
u\\
\frac{i_{L}}{G} \\
z
\end{array} \right]
\end{eqnarray}

with  the definitions $Q_{0}=\frac{C \omega_{0}}{G}$ and $i_{rn}=\frac{i_r}{C \omega_{0}}$, 
$\gamma=\frac{1}{T \omega_{0}}$ where $\omega_{0}$ is the oscillation frequency, we have:\\

\begin{eqnarray}
{\dot{X}}_{1} &=& - \frac{c X_{1} ( X_{3} -1) +X_{2}}{Q_{0}}  +i_{rn}\\
{\dot{X}}_{2} &=&  Q_{0} X_{1} \\
{\dot{X}}_{3} &=& -\gamma ( X_{3}- X_{1}^2)
\end{eqnarray}

\subsection{Clapp oscillator}

The parameters used are: \\
$$
\begin{array}{|c|c|c|c|}
\hline
R_{s} = 0  \Omega &  R_{E} = 500 \Omega  &  R_{A}  = 1000 \Omega & R_{L}  = 1000 \Omega  \\
\hline
C_{A} =100 pF & C_{E}=100 pF      &    C_{TE} =3.5 nF   & C_{TA} =7 nF  \\
\hline
\end{array}
$$

In addition, we have:   $ C_{s} =25 \hspace{1mm} pF  \mbox{ and } L_{s} =1 \hspace{1mm} mH $.\\

{\bf State-Space Equations of Clapp oscillator}:\\

\begin{eqnarray}
\frac{dV_{CE}}{dt} & = & \frac{1}{C_{E}} [ i_{e} + ( \frac{V_{CTE}-V_{CE}}{R_{E}}) ]\\
\frac{dV_{CTE}}{dt} & = & \frac{1}{C_{TE}} [-i_{p} - (\frac{V_{CTE}-V_{CE}}{R_{E}})]\\
\frac{dV_{CTA}}{dt} & = & \frac{1}{C_{TA}} [i_{p} - (\frac{V_{CTA}-V_{CA}}{R_{A}})]\\
\frac{dV_{CA}}{dt} & = & \frac{1}{C_{A}} [i_{q} +
 (\frac{V_{CTA}-V_{CA}}{R_{A}}) - \frac{V_{CA}}{R_{L}} ]\\
\frac{di_{p}}{dt} & = & j_{p} \nonumber\\
\end{eqnarray}

whereas: 

\begin{eqnarray}
&&\frac{dj_{p}}{dt}= -\frac{R_{s}j_{p}}{L_{s}} + (\frac{V_{CE}-V_{CTE}}{L_{s}R_{E}C_{TE}}) 
+ (\frac {V_{CTA}-V_{CA}}{L_{s}R_{A}C_{TA}}) \nonumber\\
&&\hspace{2cm}\mbox{}- i_{p} [ \frac{1}{L_{s}C_{TA}} + \frac{1}{L_{s}C_{s}} 
 + \frac{1}{L_{s}C_{TE}}]  
\end{eqnarray}
 
where:\\
\begin{equation}
i_{e}=A_{1} (e^{\frac{-V_{CE}}{A_{2}}}-1)
\end{equation}

and:\\
$$
i_{q} = \left[ \begin{array}{c}
-B_{1}B_{2}   \hspace{2.5cm}  -V_{CA} \le -B_{2} \hspace{2mm} \mbox{ or } V_{CA} \ge B_{2}\\
-B_{1}V_{CA}   \hspace{3.5cm}   -B_{2} <  V_{CA} < B_{2}\\
B_{1}B_{2}    \hspace{2.5cm}  -V_{CA} \ge B_{2} \hspace{2mm} \mbox{ or } V_{CA} \le -B_{2}
\end{array}
\right.
$$

\subsection{Colpitts oscillator}

$$
\begin{array}{|c|c|c|}
\hline                             
R_{1} = 350 \Omega   &    R_{2} = 110 k\Omega    &   R_{L}  = 500 \Omega \\
\hline
L_{1} =10 mH  &  L_{2} =30 nH   &    C_{1}=10 pF      \\ 
\hline
C_{2}=940 pF       &    C_{3} =2.7 nF         & C_{4} =1.5 nF \\ 
\hline
\end{array}
$$

For the transistor:\\

$$
\begin{array}{ |c| c| c|}
\hline                             
I_{s} = 2.7 \hspace{1mm} 10^{-16} A &    \beta_{I} = 5.5  &  \beta_{N}  = 140 \\
\hline
U_{A} =15 V  &   U_{B} =4.3 V &    V_{0}=12V \\    
\hline
\end{array}
$$

{\bf State-Space Equations of Coplitts oscillator}:\\

\begin{eqnarray}
\frac{di_{1}}{dt} & = & \frac{1}{L_{1}} ( V_{0} - V_{1} -R_{1}i_{1}) \\
\frac{di_{2}}{dt} & = & \frac{1}{L_{2}} ( V_{1} - V_{2}) \\
\frac{dV_{1}}{dt} & = & \frac{1}{C_{1}} [ i_{1} - i_{2}- i_{C} +( \frac{V_{4}-V_{1}}{R_{L}}) ]\\
\frac{dV_{2}}{dt} & = & \frac{1}{C_{2}} [ i_{2} - i_{B} + ( \frac{V_{0}-V_{2} -V_{3}}{R_{2}}) ]\\
\frac{dV_{3}}{dt} & = & \frac{1}{C_{3}} [ - i_{B} + ( \frac{V_{0}-V_{2} -V_{3}}{R_{2}}) ]\\
\frac{dV_{4}}{dt} & = & (\frac{V_{1}-V_{4}}{R_{L}C_{4}}) 
\end{eqnarray}

Where the transistor currents are given by:

\begin{eqnarray}
i_{C} & = & i_{CE} - i_{BC} \\
i_{B} & = & i_{BE} + i_{BC} \\
i_{CE} & = &  \frac{ I_{s} }{ ( \frac{ Q_{b} } { Q_{b0} } ) } (e^{\frac{u_{BE}}{u_{T}}}- e^{\frac{u_{BC}}{u_{T}}}) \\
i_{BE} & = & \frac{I_{s}}{ \beta_{N} } (e^{\frac{u_{BE}}{u_{T}}}- 1) \\
i_{BC} & = & \frac{I_{s}}{ \beta_{I} } (e^{\frac{u_{BC}}{u_{T}}}- 1) 
\end{eqnarray}

Moreover, we have the additional relations:

\begin{eqnarray}
u_{BE} & = & V_{2} + V_{3} \\
u_{BC} & = & V_{2} + V_{3}  - V_{1} \\
\frac{Q_{b}}{Q_{b0}} & =& 1 + \frac{u_{BE}}{u_{b}} + \frac{u_{BC}}{u_{a}} \nonumber \\
\end{eqnarray}
\begin{eqnarray}
&& i_{C}  = I_{s} \frac{(e^{\frac{V_{2} + V_{3}}{u_{T}}}- e^{\frac{ V_{2} + V_{3} - V_{1} }{u_{T}}})}{( 1 + \frac{V_{2} + V_{3}}{u_{b}} + \frac{V_{2} + V_{3} - V_{1}}{u_{a}} )} \nonumber \\
&&\hspace{1.5cm}\mbox{}  - \frac{I_{s}}{ \beta_{I} } (e^{\frac{V_{2} + V_{3} - V_{1}}{u_{T}}}- 1)  \nonumber \\
\end{eqnarray}

\begin{eqnarray}
i_{B} & = & \frac{ I_{s} }{ \beta_{N} } ( e^{ \frac{ V_{2}+V_{3}}{u_{T}}}- 1) + \frac{ I_{s} }{ \beta_{I} } ( e^{ \frac{ V_{2}+ V_{3}-V_{1} }{ u_{T} } } - 1)
\end{eqnarray}

Define:

\begin{eqnarray}
\left[ \begin{array}{c}
X_{1} \\
X_{2}  \\
X_{3} \\
X_{4} \\
X_{5}  \\
X_{6} 
\end{array} \right] \stackrel{Def.}{=} \left[ \begin{array}{c}
i_{1} \\
i_{2}  \\
V_{1} \\
V_{2} \\
V_{3}  \\
V_{4} 
\end{array} \right]
\end{eqnarray}

Rewrite Equations in state-space form:

\begin{eqnarray}
{\dot{X}}_{1} & = &   \frac{1}{L_{1}} ( V_{0} - X_{3} -R_{1}X_{1})\\
{\dot{X}}_{2} & = &  \frac{1}{L_{2}} ( X_{3} - X_{4}) \nonumber \\
\end{eqnarray}
\begin{eqnarray}
&& {\dot{X}}_{3} =   \frac{1}{C_{1}} [ X_{1} - X_{2} + \frac{1}{R_{L}} (X_{6} -X_{3}) +
 \frac{I_{s}}{ \beta_{I} } (e^{\frac{X_{4}+X_{5} -X_{3} }{u_{T}}}- 1) \nonumber \\
&&\hspace{0.5cm}\mbox{} - \frac{I_{s}}{(1 + \frac{X_{4} + X_{5}}{u_{b}} + \frac{X_{4} + X_{5} - X_{3}}{u_{a}} )}
(e^{\frac{X_{4} + X_{5}}{u_{T}}}- e^{\frac{ X_{4} + X_{5} - X_{3} }{u_{T}}} ) ] \nonumber \\
\end{eqnarray}
\begin{eqnarray}
&&{\dot{X}}_{4} =   \frac{1}{ C_{2} } [ X_{2} + 
\frac{1}{R_{2}} ( V_{0} - X_{4} -X_{5} )  
- \frac{I_{s}}{ \beta_{N} } (e^{ \frac{X_{4}
 + X_{5}}{u_{T}}}- 1)  \nonumber \\
&&\hspace{0.5cm}\mbox{} - \frac{I_{s}}{ \beta_{I} }
 (e^{ \frac{ X_{4} + X_{5} - X_{3} }{u_{T} } }- 1)   ] \nonumber\\
\end{eqnarray}
\begin{eqnarray}
&&{\dot{X}}_{5} =   \frac{1}{C_{3}} [ \frac{1}{R_{2}}
 ( V_{0} - X_{4} -X_{5})  - \frac{I_{s}}{ \beta_{N} }
 (e^{\frac{X_{4} + X_{5}}{u_{T}}}- 1) \nonumber\\ 
&&\hspace{0.5cm}\mbox{} - \frac{I_{s}}{ \beta_{I} }
 (e^{\frac{ X_{4} + X_{5} - X_{3} }{u_{T}}}- 1)   ] \nonumber\\
\end{eqnarray}
\begin{eqnarray}
{\dot{X}}_{6} & = &   \frac{1}{R_{L} C_{4}} ( X_{3} - X_{6})
\end{eqnarray}

\subsection{Model I}

$$
\begin{array}{|c|c|c|}
\hline                             
V_{0} = 9 V   &    R_{1} = 220 k \Omega    &   R_{2}  = 1000 \Omega \\
\hline
R_{3} =220 k \Omega &  R_{4} =2 \Omega   &    L =10 \mu H      \\ 
\hline
C_{1}=0.47 \mu F       &    C_{2} =200 pF         & C_{3} =200 pF \\ 
\hline
\end{array}
$$

For the transistor:\\

$$
\begin{array}{ |c| c| c| c| c|}
\hline                             
I_{s} = 2.7 \hspace{1mm} 10^{-16} A &  U_{A} =15 V  &   U_{B} =4.3 V \\
\hline
Gain &   \beta_{I} = 5.5  &  \beta_{N}  = 140 \\
\hline
\end{array}
$$

{\bf State-Space Equations of Model I oscillator}:\\

\begin{eqnarray}
\frac{di_{4}}{dt} & = & \frac{1}{L} ( V_{6} + V_{7} -R_{4}i_{4} - V_{5}) \\
\frac{dV_{5}}{dt} & = & \frac{i_{4}}{C_{1}} \nonumber \\
\end{eqnarray}
\begin{eqnarray}
&&\frac{dV_{6}}{dt}  =  \frac{1}{C_{2}} [  ( \frac{V_{0}-V_{6} -V_{7}}{R_{1}})  - ( \frac{V_{6} +V_{7}}{R_{3}})  \nonumber \\
&&\hspace{2.5cm}\mbox{}   -i_{4} - i_{B} ] \nonumber \\
\end{eqnarray}
\begin{eqnarray}
&& \frac{dV_{7}}{dt} = \frac{1}{C_{3}} [  ( \frac{V_{0}-V_{6} -V_{7}}{R_{1}})  - ( \frac{V_{6} +V_{7}}{R_{3}}) \nonumber \\
&&\hspace{0.5cm}\mbox{} -i_{4} - i_{B}  +i_{E} -  \frac{V_{7}}{R_{2}} ] \nonumber \\
\end{eqnarray}

The currents and transistor voltages $u_{BE}$ and $u_{BC}$ are given by:

\begin{eqnarray}
i_{E} & = & i_{B} + i_{C} \\
u_{BE} & = & V_{6} \\
u_{BC} & = & V_{6} + V_{7} - V_{0} \nonumber \\
\end{eqnarray}

Therefore,  $i_{B} \mbox{ and } i_{C}$ are given by:

\begin{eqnarray}
i_{B} & = & \frac{ I_{s} }{ \beta_{N} } ( e^{ \frac{V_{6}}{u_{T}}}- 1) + \frac{ I_{s} }{ \beta_{I} } ( e^{ \frac{ V_{6}+ V_{7}-V_{0} }{ u_{T} } } - 1) \nonumber \\
\end{eqnarray}
\begin{eqnarray}
&& i_{C} =  \frac{I_{s}}{( 1 + \frac{V_{6}}{u_{b}} + \frac{V_{6} + V_{7} - V_{0}}{u_{a}} )} (e^{\frac{V_{6}}{u_{T}}}- e^{\frac{ V_{6} + V_{7} - V_{0} }{u_{T}}}) \nonumber \\
&&\hspace{2.5cm}\mbox{} - \frac{I_{s}}{ \beta_{I} } (e^{\frac{V_{6} + V_{7} - V_{0}}{u_{T}}}- 1) \nonumber \\
\end{eqnarray}

Define:

\begin{eqnarray}
\left[ \begin{array}{c}
X_{1} \\
X_{2}  \\
X_{3} \\
X_{4} 
\end{array} \right] \stackrel{Def.}{=} \left[ \begin{array}{c}
i_{4} \\
V_{3} \\
V_{6} \\
V_{7} 
\end{array} \right]
\end{eqnarray}

Rewrite equations in state-space form:

\begin{eqnarray}
{\dot{X}}_{1} & = &   \frac{1}{L} ( X_{3} + X_{4} - X_{2} -R_{4}X_{1}) \\
{\dot{X}}_{2} & = &  \frac{X_{1}}{C_{1}} \nonumber \\
\end{eqnarray}
\begin{eqnarray}
&&{\dot{X}}_{3} = \frac{1}{C_{2}} [ -X_{1} + \frac{1}{R_{1}} (V_{0} - X_{3} - X_{4}) - \frac{1}{R_{3}} (X_{3} + X_{4}) \nonumber \\
&&\hspace{0.5cm}\mbox{} - \frac{I_{s}}{ \beta_{N} } (e^{\frac{X_{3}}{u_{T}}}- 1)- \frac{I_{s}}{ \beta_{I} } (e^{\frac{X_{3} + X_{4} - V_{0}}{u_{T}}}- 1)] \nonumber \\
\end{eqnarray}
\begin{eqnarray}
&&{\dot{X}}_{4} =   \frac{1}{C_{3}} [ \frac{1}{R_{1}} ( V_{0} - X_{3} - X_{4})  - X_{1 } - \frac{1}{R_{3}} ( X_{3} + X_{4}) \nonumber \\
&&\hspace{1.5cm}\mbox{} - \frac{X_{4}}{R_{2}} - \frac{I_{s}}{ \beta_{I} } (e^{\frac{X_{3} + X_{4} - V_{0}}{u_{T}}}- 1) \nonumber \\
&&\hspace{0.5cm}\mbox{} + \frac{I_{s}}{( 1 + \frac{X_{3}}{u_{b}} + \frac{X_{3} + X_{4}  - V_{0}}{u_{a}} )} (e^{\frac{X_{3}}{u_{T}}}- e^{\frac{ X_{3} + X_{4} - V_{0} }{u_{T}}} ) ] \nonumber \\
\end{eqnarray}

\subsection{Model II}

$$
\begin{array}{|c|c|c|c|}
\hline                             
 R_{1} = 1000 \Omega    &   R_{2}  = 82 k \Omega  &    R_{3}  = 680 \Omega &    L =100 nH      \\
\hline
C_{1}=33 pF       &    C_{2} =33 pF         & C_{3} =10 pF   & C_{4} = 0.1 \mu F \\ 
\hline
\end{array}
$$

For the transistor:\\

$$
\begin{array}{ |c| c| c|}
\hline                             
I_{s} = 2.7 \hspace{1mm} 10^{-16} A &   U_{A} =15 V  &   U_{B} =4.3 V \\
\hline
Gain & \beta_{I} = 5.5  &  \beta_{N}  = 140  \\
\hline
\end{array}
$$

{\bf State-Space Equations of Model II oscillator}:\\
The transistor voltages  and   are given by:

\begin{eqnarray}
&& \frac{dV_{1}}{dt} = \frac{1}{C_{1}} [  ( \frac{V_{0}-V_{1}}{R_{1}}) -i_{C} - ( \frac{V_{1} -V_{2}}{R_{2}}) \nonumber \\
&&\hspace{2.5cm}\mbox{} - ( \frac{V_{1} -V_{2} -V_{5}}{R_{3}}) - i_{6} ] \nonumber \\
\end{eqnarray}
\begin{eqnarray}
&& \frac{dV_{2}}{dt} = \frac{1}{C_{2}} [  ( \frac{V_{1}-V_{2}}{R_{2}})  + ( \frac{V_{1} -V_{2} -V_{5}}{R_{3}})  \nonumber \\
&&\hspace{3.5cm}\mbox{} +i_{6}- i_{B} ] \nonumber \\
\end{eqnarray}
\begin{eqnarray}
\frac{dV_{3}}{dt} & = & \frac{i_{6}}{C_{3}} \\
\frac{di_{6}}{dt} & = & \frac{1}{L} ( V_{1} - V_{2} - V_{5}) \\
\frac{dV_{5}}{dt} & = & \frac{1}{C_{4}}  ( \frac{V_{1}-V_{2} -V_{5} }{R_{3}} ) \nonumber \\
\end{eqnarray}

The currents and transistor voltages $u_{BE}$ and $u_{BC}$ are given by:

\begin{eqnarray}
u_{BE} & = & V_{2} \\
u_{BC} & = & V_{2}  - V_{1} \nonumber \\
\end{eqnarray}

Therefore,  $i_{B} \mbox{ and } i_{C}$ are given by:

\begin{eqnarray}
i_{B} & = & \frac{ I_{s} }{ \beta_{N} } ( e^{ \frac{V_{2}}{u_{T}}}- 1) + \frac{ I_{s} }{ \beta_{I} }
 ( e^{ \frac{ V_{2}-V_{1} }{ u_{T} } } - 1) \nonumber \\
\end{eqnarray}
\begin{eqnarray}
&& i_{C} = \frac{I_{s}}{( 1 + \frac{V_{2}}{u_{b}} + \frac{V_{2} - V_{1}}{u_{a}} )} (e^{\frac{V_{2}}{u_{T}}}- e^{\frac{ V_{2} - V_{1} }{u_{T}}})  \nonumber \\
&&\hspace{2cm}\mbox{} - \frac{I_{s}}{ \beta_{I} } (e^{\frac{V_{2} - V_{1}}{u_{T}}}- 1)  \nonumber \\
\end{eqnarray}

Define:

\begin{eqnarray}
\left[ \begin{array}{c}
X_{1} \\
X_{2}  \\
X_{3} \\
X_{4} \\
X_{5}
\end{array} \right] \stackrel{Def.}{=} \left[ \begin{array}{c}
V_{1} \\
V_{2} \\
V_{3} \\
i_{6} \\
V_{5} 
\end{array} \right]
\end{eqnarray}

Rewrite equations in state-space form:

\begin{eqnarray}
&& {\dot{X}}_{1} =   \frac{1}{C_{1}} [ \frac{1}{R_{1}} (V_{0} - X_{1 }) - \frac{1}{R_{2}} ( X_{1} - X_{2}) - X_{4} \nonumber \\
&&\hspace{1.5cm}\mbox{} - \frac{1}{R_{3}} ( X_{1} - X_{2}  - X_{5} ) + 
\frac{I_{s}}{ \beta_{I} } (e^{\frac{X_{2} - X_{1}}{u_{T}}}- 1)  \nonumber \\
&&\hspace{1.5cm}\mbox{} - \frac{I_{s}}{( 1 + \frac{X_{2}}{u_{b}} + \frac{X_{2} - X_{1}}{u_{a}} )} (e^{\frac{X_{2}}{u_{T}}}- e^{\frac{ X_{2} - X_{1} }{u_{T}}} ) ]  \nonumber \\
\end{eqnarray}
\begin{eqnarray}
&& {\dot{X}}_{2}  =  \frac{1}{C_{2}} [ X_{4} + \frac{1}{R_{2}} (X_{1} - X_{2}) + \frac{1}{R_{3}} ( X_{1} - X_{2} - X_{5}) \nonumber \\ 
&&\hspace{1.5cm}\mbox{} - \frac{I_{s}}{ \beta_{N} } (e^{\frac{X_{2}}{u_{T}}}- 1)- \frac{I_{s}}{ \beta_{I} } (e^{\frac{X_{2} - X_{1}}{u_{T}}}- 1)]  \nonumber \\
\end{eqnarray}
\begin{eqnarray}
{\dot{X}}_{3} & = &  \frac{X_{4}}{C_{3}} \\
{\dot{X}}_{4} & = &   \frac{1}{L} ( X_{1} - X_{2} - X_{3} )\\
{\dot{X}}_{5} & = &   \frac{1}{ C_{4} R_{3}  } ( X_{1} - X_{2} - X_{5} )  \nonumber \\
\end{eqnarray}

\centerline{\Large\bf FIGURE CAPTIONS}

\begin{figure}[htbp]
\begin{center}
\scalebox{0.5}{\includegraphics*{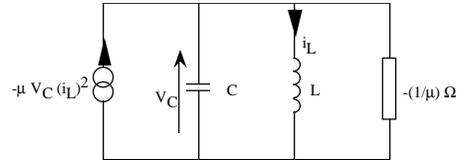}}
\end{center}
\caption{Van der Pol oscillator circuit}
\label{fig1}
\end{figure}

\begin{figure}[htbp]
\begin{center}
\scalebox{0.5}{\includegraphics*{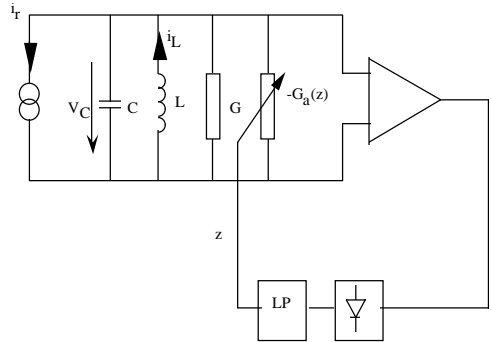}}
\end{center}
\caption{Amplitude controlled Van der Pol oscillator circuit}
\label{fig2}
\end{figure}

\begin{figure}[htbp]
\begin{center}
\scalebox{0.5}{\includegraphics*{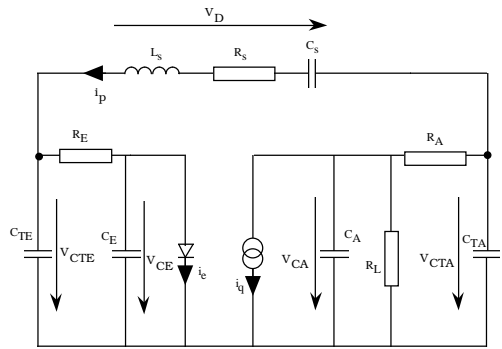}}
\end{center}
\caption{Clapp oscillator circuit}
\label{fig3}
\end{figure}

\begin{figure}[htbp]
\begin{center}
\scalebox{0.5}{\includegraphics*{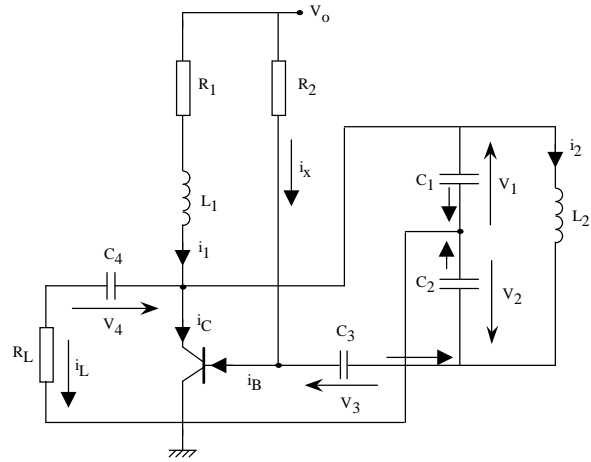}}
\end{center}
\caption{Colpitts oscillator circuit}
\label{fig4}
\end{figure}

\begin{figure}[htbp]
\begin{center}
\scalebox{0.5}{\includegraphics*{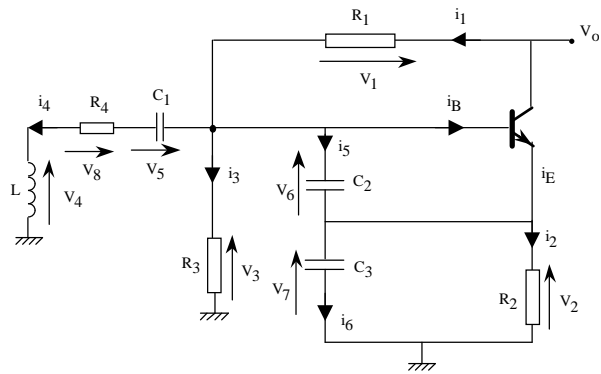}}
\end{center}
\caption{Model 1 oscillator circuit}
\label{fig5}
\end{figure}

\begin{figure}[htbp]
\begin{center}
\scalebox{0.5}{\includegraphics*{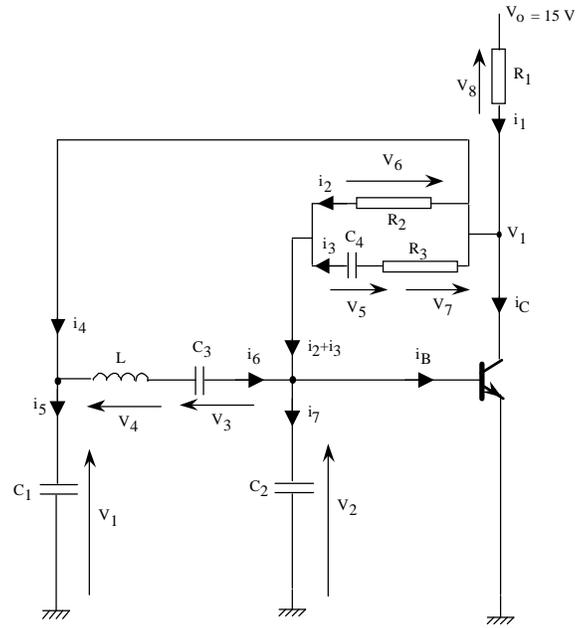}}
\end{center}
\caption{Model II oscillator circuit}
\label{fig6}
\end{figure}

\end{document}